\documentclass{moriond}
\usepackage{amsmath}
\usepackage{caption}
\newcommand{\Photo}{}
\def\mevcc{\mathrm{\,Me\kern -0.1em V\!/}c^2}
\def\gevcc{\mathrm{\,Ge\kern -0.1em V\!/}c^2}
\def\tev{\mathrm{\,Te\kern -0.1em V}}
\begin{document}
\vspace*{4cm}
\title{Electroweak scale physics \& exotic searches at LHCb}
\author{O. Lupton on behalf of the LHCb collaboration}
\address{European Organization for Nuclear Research (CERN), Meyrin, Switzerland}
\maketitle\abstracts{The LHCb experiment has a broad and varied physics programme, extending far beyond its core set of
flavour physics measurements. This contribution summarises recent electroweak scale measurements and searches for exotic
states in the dimuon final state.}

\section{Introduction}
The LHCb detector is a single-arm forward spectrometer covering the pseudorapidity range 2--5 that is principally
designed for the study of $b$- and $c$-hadrons, but which is well-suited to a wide variety of electroweak scale
measurements and exotic searches that are highly complementary to other experiments at the LHC and elsewhere.
Several features of the detector that are crucial for the core flavour physics programme, such as excellent vertex and
momentum resolution, and a powerful trigger system, contribute to excellent jet tagging performance and sensitivity to
low mass exotic states.
LHCb operates at a substantially lower instantaneous luminosity than the general purpose detectors at the LHC, ATLAS
and CMS, which results in a clean, low pile-up environment in which to search for physics beyond the Standard Model
(SM).

\section{Searches for exotic dimuon resonances}
The excellent mass resolution and muon identification performance of the detector over a wide range of momenta make
searches for exotic dimuon resonances an attractive prospect in LHCb data.
Two results are presented in this section, one covering a wide mass range and another focused on the region in the
immediate vicinity of the $\Upsilon$ resonances.

\subsection{Dark photons}
Several theories of physics beyond the SM propose new particles that could explain the nature of dark matter.
Such theories typically include additional particles and dark boson-mediated forces, with a massive dark photon $A'$
being a popular feature.
These dark photons typically do not couple directly to charged SM particles, but they can gain a weak coupling to the
SM electromagnetic current via kinetic mixing.
The strength of this coupling is suppressed by a factor $\varepsilon$ with respect to the SM photon.
There is, therefore, a 2D parameter space to be probed.

\begin{figure}
  \includegraphics[width=\textwidth]{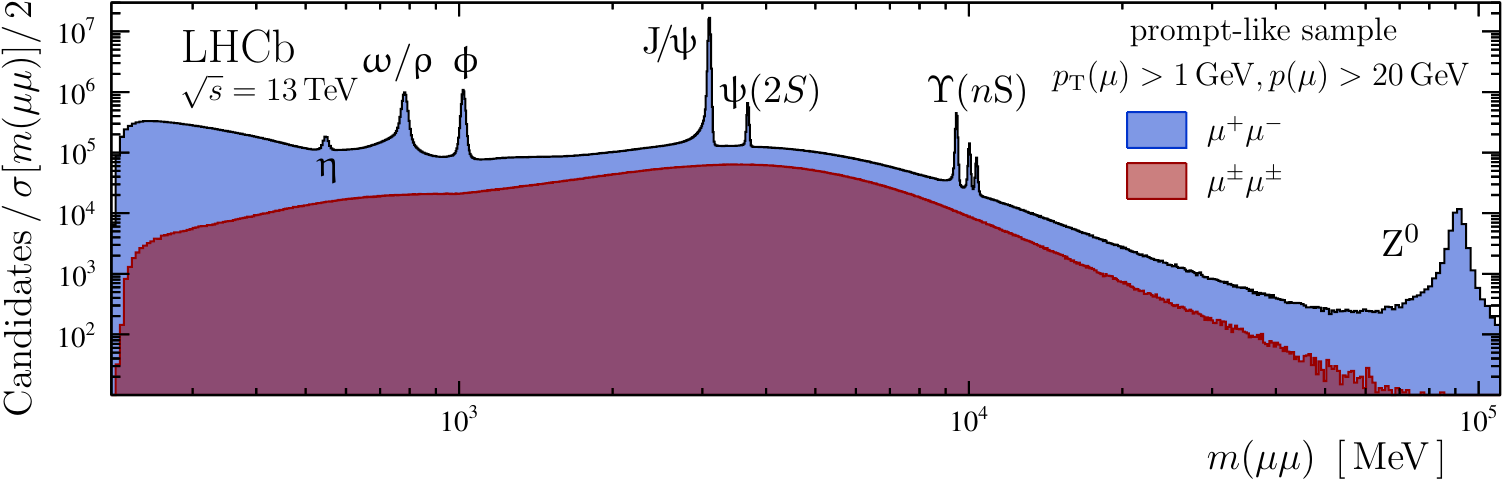}
  \caption[]{Prompt-like dimuon spectrum used in the dark photon search with various resonance peaks labelled.\cite{LHCb-PAPER-2017-038}}
  \label{fig:darkphoton_dimuon_spectrum}
\end{figure}

The LHCb measurement\cite{LHCb-PAPER-2017-038} presented here uses the dimuon final state and is based on a dataset
corresponding to an integrated luminosity of $\mathcal{L} = 1.6\mathrm{\,fb^{-1}}$ recorded at $\sqrt{s}$ = 13 TeV
during 2016. The flexible nature of the LHCb software trigger\cite{LHCb-DP-2016-001} allows the full rate of prompt,
\textit{i.e.} consistent with originating at the primary $pp$ interaction vertex, dimuon candidates to be recorded for
analysis, from the dimuon mass threshold up to the $Z^0$.
The prompt-like dimuon spectrum is shown in Fig.~\ref{fig:darkphoton_dimuon_spectrum}.
Backgrounds due to semileptonic decays of heavy-flavour hadrons, and misidentified hadrons, are determined using fits to
$\chi^2$-like variables relating to the dimuon vertex quality.
Background due to off-shell photon decays $\gamma^{\ast}\to\mu^{+}\mu^{-}$ is irreducible and used to normalise the dark
photon search in a data-driven manner.
The various well-known quarkonia peaks labelled in Fig.~\ref{fig:darkphoton_dimuon_spectrum} are excluded from the
prompt-like search, which extends up to $70\gevcc$.

A search is performed for displaced dimuon vertices in the mass range $[214, 350]\mevcc$, in this case the
background composition is somewhat different. At radii of more than $5\mathrm{\,mm}$ and low mass the background is
dominated by real photon conversions in the vertex locator material. Such conversions are vetoed using a new method
based on a large dataset of secondary hadronic interaction vertices\cite{LHCb-DP-2018-002}. At smaller radii the
background is dominated by heavy flavour decays.

\begin{figure}
  \includegraphics[width=\textwidth]{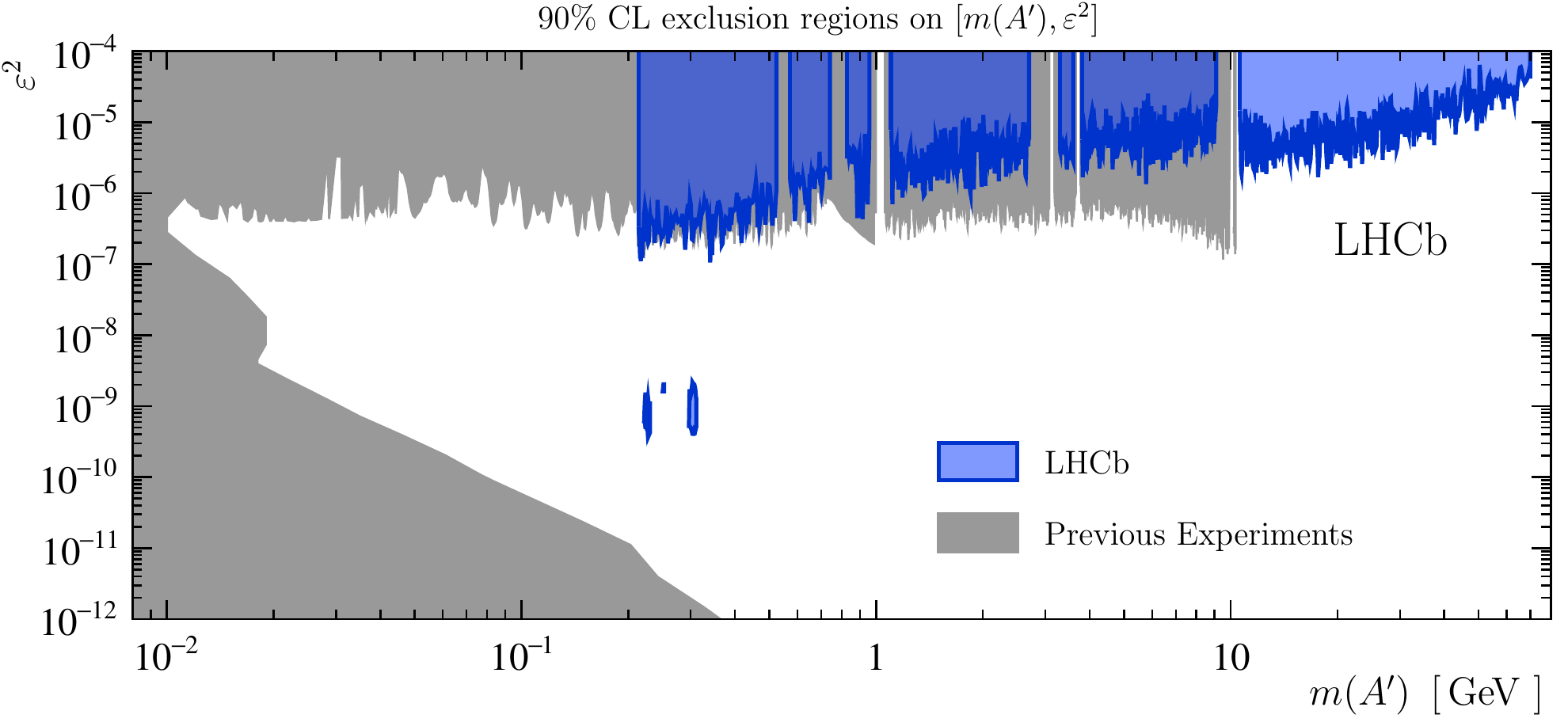}
  \caption[]{Results of the dark photon search. Both prompt-like (top) and displaced (centre) exclusions are
             shown.\cite{LHCb-PAPER-2017-038}}
  \label{fig:darkphoton_exclusions}
\end{figure}

No significant excess is found, and the limits set are shown in Fig.~\ref{fig:darkphoton_exclusions}.
The prompt-like search sets world-best limits in the mass region $[10.7, 70]\gevcc$, while the displaced search probes
a world-first region of parameter space.
Future updates to these searches are expected to take advantage of the novel all-software trigger in the LHCb experiment
upgrade for Run 3 of the LHC, and to extend to lower values of $m(A')$ by exploiting the dielectron
channel.\cite{Ilten:2015hya,Ilten:2016tkc}

\subsection{Search in vicinity of the $\boldmath\varUpsilon$ resonances}
\begin{figure}
  \begin{minipage}{0.48\textwidth}
    \includegraphics[width=\textwidth]{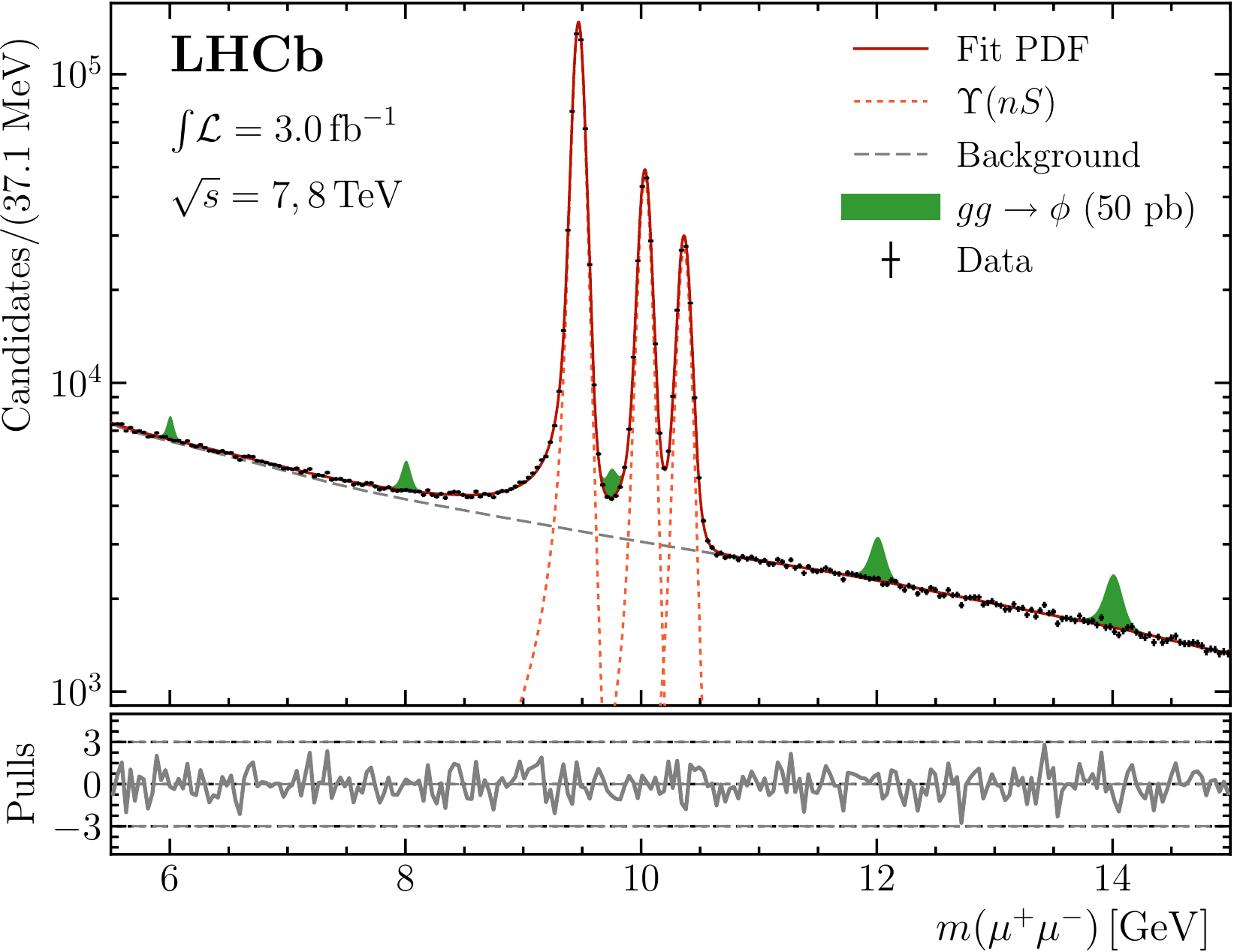}
    \captionof{figure}[]{Dimuon spectrum in the vicinity of the $\Upsilon(n\mathrm{S})$
                         resonances\cite{LHCb-PAPER-2018-008}.
                         Peaks corresponding to scalar resonances $\phi$ are also shown, where it has been assumed that
                         $\sigma(pp\to\phi)\mathcal{B}(\phi\to\mu^{+}\mu^{-})=50\,\mathrm{pb}$.\cite{LHCb-PAPER-2018-008}}
    \label{fig:upsilon_dimuon_spectrum}
  \end{minipage}
  \hfill
  \begin{minipage}{0.48\textwidth}
    \vspace{10mm}
    \includegraphics[width=\textwidth]{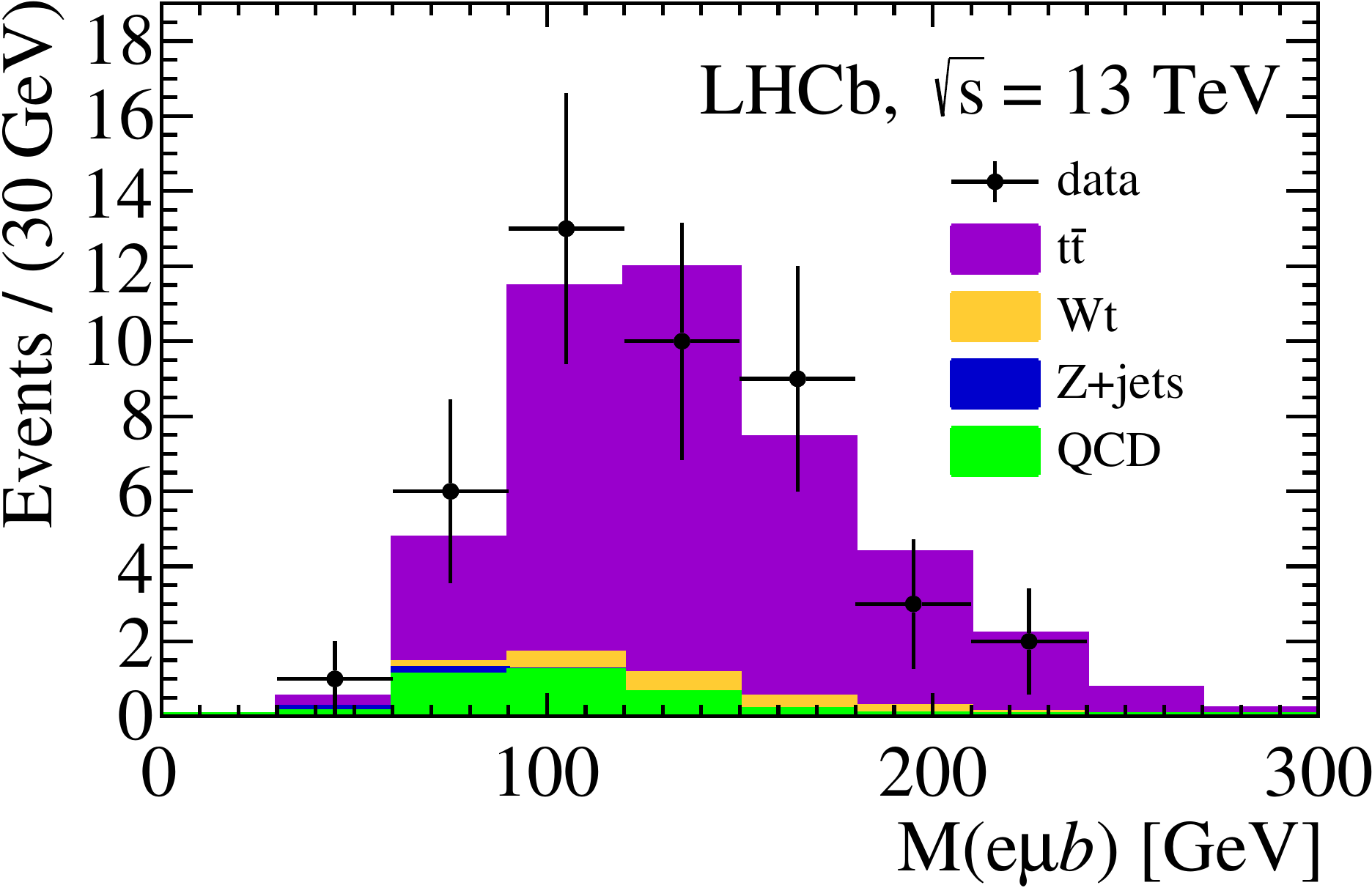}
    \captionof{figure}[]{$e\mu b$ invariant mass for all 44 selected candidates in the $t\bar{t}$
                         measurement\cite{LHCb-PAPER-2017-050}.
                         No fit is performed to this distribution, it is presented simply to illustrate the signal
                         purity.}
    \label{fig:ttbar_mass}
  \end{minipage}
\end{figure}

A dedicated search for dimuon resonances in the vicinity of the $\Upsilon(n\mathrm{S})$ resonances has also been
performed, probing closer to, and in between, these peaks than the dark photon search.\cite{LHCb-PAPER-2018-008}
The mass region that is probed in this analysis is illustrated in Fig.~\ref{fig:upsilon_dimuon_spectrum}.
No significant excess is seen, but the first limits are set in the mass range $[8.7, 11.5]\gevcc$.
Limits are set assuming that the produced resonance is either a scalar or a vector boson.

\section{\boldmath$t\bar{t}$ production}
Top quark physics forms an important part of the electroweak scale physics programme at LHCb, and is an excellent
example of an area of study where the forward acceptance of the LHCb detector has several advantages with respect to the
central region instrumented by ATLAS and CMS.
For example, top quark cross-sections can provide important constraints on the large-$x$ gluon PDF, with the forward
kinematic region being particularly sensitive.
The forward region also provides a greater fraction of events with quark-initiated production than central detectors,
enhancing the size of $t\bar{t}$ asymmetries visible at LHCb.

Carrying out such measurements at LHCb presents several challenges relating to the small acceptance, low luminosity and
lack of missing energy measurement.
Despite these challenges, $t$ and $t\bar{t}$ production\cite{LHCb-PAPER-2015-022,LHCb-PAPER-2016-038} have previously
been observed at LHCb using data recorded at $\sqrt{s} = 7\tev$ and $8\tev$.

\begin{figure}
  \begin{minipage}{0.56\textwidth}
    \vspace{5mm}
    \includegraphics[width=\textwidth]{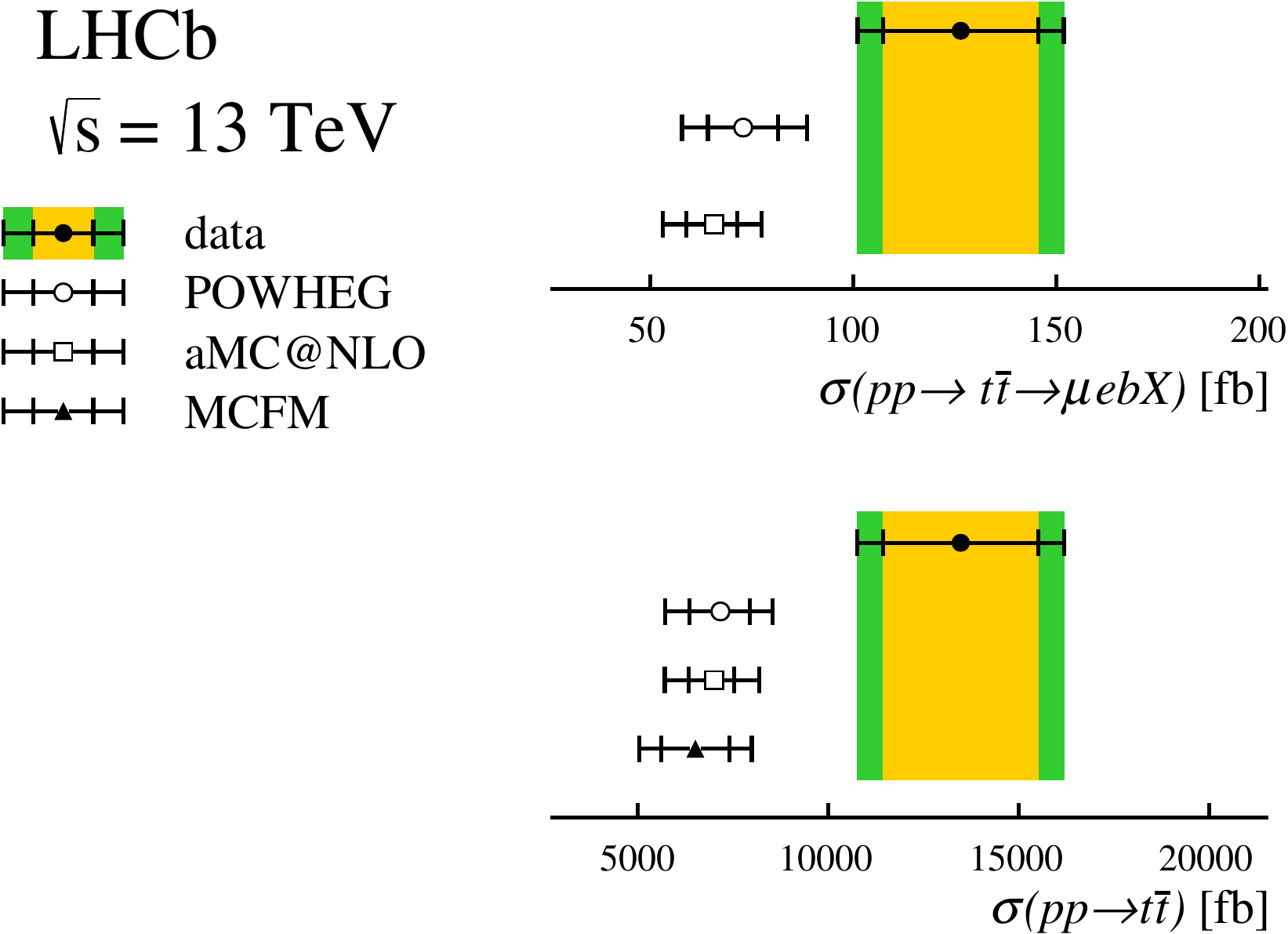}
    \captionof{figure}[]{Fiducial (top) and extrapolated (bottom) $t\bar{t}$ cross sections, compared to SM predictions
                         at next-to-leading order\cite{LHCb-PAPER-2017-050}.}
    \label{fig:ttbar_xsecs}
  \end{minipage}
  \hfill
  \begin{minipage}{0.4\textwidth}
    \includegraphics[width=\textwidth]{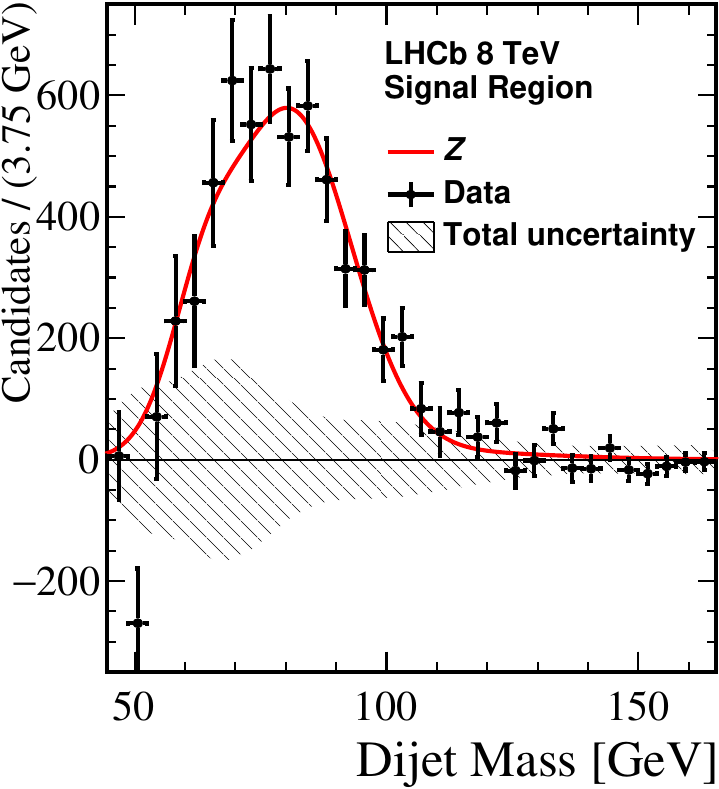}
    \captionof{figure}[]{Background subtracted dijet mass spectrum showing the $Z^0\to b\bar{b}$ signal\cite{LHCb-PAPER-2017-024}.}
    \label{fig:bbbar_peak}
  \end{minipage}
\end{figure}

The step to $\sqrt{s} = 13\tev$ for Run 2 of the LHC has increased the LHCb-visible cross sections for top quark
processes by an order of magnitude, bringing new channels into statistical reach.
The new result presented here is a measurement of $t\bar{t}$ production at $\sqrt{s} = 13\tev$ using the $e\mu b$ final
state\cite{LHCb-PAPER-2017-050} and an integrated luminosity of $\mathcal{L} = 2\mathrm{\,fb^{-1}}$.
This is a very pure final state, as the second lepton suppresses $W + b\bar{b}$ production and the opposite-flavour
leptons suppress $Z^0 + b\bar{b}$.
The signal purity is illustrated in Fig.~\ref{fig:ttbar_mass}, and the fiducial cross section is calculated as
\begin{equation}
  \sigma_{t\bar{t}} = \frac{N - N_{\mathrm{bkg}}}{\mathcal{L}\cdot\epsilon}\cdot\mathcal{F}_{\mathrm{res}} =126 \pm 19\mathrm{\,(stat)} \pm 16\mathrm{\,(syst)} \pm 5\mathrm{\,(lumi)}\,\mathrm{fb^{-1}}. \\
  \label{eq:xsec}
\end{equation}
Figure~\ref{fig:ttbar_xsecs} compares this result, and an extrapolation to the full cross section, with SM predictions
at next-to-leading order. The measured cross sections are compatible with these predictions.

\section{\boldmath$Z^0\to b\bar{b}$}
The final measurement discussed in these proceedings is the first observation of $Z^0\to b\bar{b}$ production in the
forward region\cite{LHCb-PAPER-2017-024}.
This measurement is an important validation of the LHCb jet reconstruction and $b$-tagging performance.
Two $b$-tagged jets are reconstructed, with a third balancing jet also reconstructed to help control the QCD background
and define signal and control regions using a multivariate technique.
The background-subtracted signal distribution is shown in Fig.~\ref{fig:bbbar_peak}.
The signal is observed with a statistical significance of $6\sigma$ and the measured cross section is found to be
compatible with SM predictions at next-to-leading order.

\section{Conclusions}
LHCb has a broad and exciting programme of electroweak and exotic measurements, which are typically still statistically
limited and will benefit significantly from further Run 2 updates and, crucially, from the LHCb upgrade ahead of Run 3
of the LHC.

\section*{References} 


\begin{thebibliography}{99}
\bibitem{LHCb-PAPER-2017-038} 
LHCb collaboration,
\emph{{Search for dark photons produced in 13 TeV $pp$ collisions}},
\href{http://dx.doi.org/10.1103/PhysRevLett.120.061801}{Phys.\ Rev.\ Lett.\  \textbf{120} (2018) 061801},
\href{http://arxiv.org/abs/1710.02867}{{\normalfont\ttfamily arXiv:1710.02867}}

\bibitem{LHCb-DP-2016-001}
R.~Aaij,
\emph{{Tesla: an application for real-time data analysis in High Energy Physics}},
\href{http://dx.doi.org/10.1016/j.cpc.2016.07.022}{Comput.\ Phys.\ Commun.\  \textbf{208} (2016) 35},
\href{http://arxiv.org/abs/1604.05596}{{\normalfont\ttfamily arXiv:1604.05596}}

\bibitem{LHCb-DP-2018-002}
M.~Alexander {\em et~al.},
\emph{{Mapping the material in the LHCb vertex locator using secondary hadronic interactions}},
\href{http://arxiv.org/abs/1803.07466}{{\normalfont\ttfamily arXiv:1803.07466}}

\bibitem{Ilten:2015hya}
P.~Ilten, J.~Thaler, M.~Williams, and W.~Xue,
\emph{{Dark photons from charm mesons at LHCb}}
\href{http://dx.doi.org/10.1103/PhysRevD.92.115017}{Phys.\ Rev.\ \textbf{D92} (2015) 115017},
\href{http://arxiv.org/abs/1509.06765}{{\normalfont\ttfamily arXiv:1509.06765}}

\bibitem{Ilten:2016tkc}
P.~Ilten {\em et~al.},
\emph{{Proposed inclusive dark photon search at LHCb}},
\href{http://dx.doi.org/10.1103/PhysRevLett.116.251803}{Phys.\ Rev.\ Lett.\  \textbf{116} (2016) 251803},
\href{http://arxiv.org/abs/1603.08926}{{\normalfont\ttfamily arXiv:1603.08926}}

\bibitem{LHCb-PAPER-2018-008}
LHCb collaboration,
\emph{{Search for a dimuon resonance in the $\varUpsilon$ mass region}},
\href{http://arxiv.org/abs/1805.09820}{{\normalfont\ttfamily arXiv:1805.09820}}, submitted to JHEP

\bibitem{LHCb-PAPER-2015-022}
LHCb collaboration,
\emph{{First observation of top quark production in the forward region}},
\href{http://dx.doi.org/10.1103/PhysRevLett.115.112001}{Phys.\ Rev.\ Lett.\  \textbf{115} (2015) 112001},
\href{http://arxiv.org/abs/1506.00903}{{\normalfont\ttfamily arXiv:1506.00903}}

\bibitem{LHCb-PAPER-2016-038}
LHCb collaboration,
\emph{{Measurement of the $t\bar{t}$, $W + b\bar{b}$ and $W + c\bar{c}$ production cross sections in $pp$ collisions at $\sqrt{s}$ = 8 TeV}},
\href{http://dx.doi.org/10.1016/j.physletb.2017.01.044}{Phys.\ Lett.\ \textbf{B767} (2017) 110},
\href{http://arxiv.org/abs/1610.08142}{{\normalfont\ttfamily arXiv:1610.08142}}

\bibitem{LHCb-PAPER-2017-050} 
LHCb collaboration,
\emph{{Measurement of forward top pair production in the dilepton channel in $pp$ collisions at $\sqrt{s}$ = 13 TeV}},
\href{http://arxiv.org/abs/1803.05188}{{\normalfont\ttfamily arXiv:1803.05188}},
submitted to JHEP

\bibitem{LHCb-PAPER-2017-024}
LHCb collaboration,
\emph{{First observation of forward $Z\to b\bar{b}$ production in $pp$ collisions at $\sqrt{s}$ = 8 TeV}},
\href{http://dx.doi.org/10.1016/j.physletb.2017.11.066}{Phys.\ Lett.\ \textbf{B776} (2017) 430},
\href{http://arxiv.org/abs/1709.03458}{{\normalfont\ttfamily arXiv:1709.03458}}

\end{thebibliography}
\end{document}